\documentclass[manuscript,screen,authorversion]{acmart}

\copyrightyear{2024}
\acmYear{2024}
\begin{document}

\title{Empathy and the Right to Be an Exception: What LLMs Can and Cannot Do}

\author{William Kidder}
\email{william.g.kidder@gmail.com}
\affiliation{%
  \institution{Status Labs}
  \country{USA}
}

\author{Jason D'Cruz}
\email{jdcruz@albany.edu}
\orcid{0000-0001-9839-7752}
\affiliation{%
  \institution{University at Albany, State University of New York}
  \streetaddress{1400 Washington Avenue}
  \city{Albany}
  \state{New York}
  \country{USA}
  \postcode{12222}
}

\author{Kush R. Varshney}
\email{krvarshn@us.ibm.com}
\orcid{0000-0002-7376-5536}
\affiliation{%
  \institution{IBM Research -- Thomas J. Watson Research Center}
  \streetaddress{1101 Kitchawan Road}
  \city{Yorktown Heights}
  \state{New York}
  \country{USA}
  \postcode{10598}
}


\begin{abstract}

Advances in the performance of large language models (LLMs) have led some researchers to propose the emergence of theory of mind (ToM) in artificial intelligence (AI). LLMs can attribute beliefs, desires, intentions, and emotions, and they will improve in their accuracy. Rather than employing the characteristically human method of empathy, they learn to attribute mental states by recognizing linguistic patterns in a dataset that typically do not include that individual. We ask whether LLMs’ inability to empathize precludes them from honoring an individual’s right to be an exception, that is, from making assessments of character and predictions of behavior that reflect appropriate sensitivity to a person's individuality. Can LLMs seriously consider an individual’s claim that their case is different based on internal mental states like beliefs, desires, and intentions, or are they limited to judging that case based on its similarities to others? We propose that the method of empathy has special significance for honoring the right to be an exception that is distinct from the value of predictive accuracy, at which LLMs excel. We conclude by considering whether using empathy to consider exceptional cases has intrinsic or merely practical value and we introduce conceptual and empirical avenues for advancing this investigation.
\end{abstract}

\begin{CCSXML}
<ccs2012>
   <concept>
       <concept_id>10010147.10010178.10010216</concept_id>
       <concept_desc>Computing methodologies~Philosophical/theoretical foundations of artificial intelligence</concept_desc>
       <concept_significance>500</concept_significance>
       </concept>
 </ccs2012>
\end{CCSXML}

\ccsdesc[500]{Computing methodologies~Philosophical/theoretical foundations of artificial intelligence}
\keywords{LLM, Empathy, Artificial Intelligence, Moral Psychology, Theory of Mind}


\maketitle

\section{Introduction}
\label{sec:intro}

\emph{I see why this looks bad. But \emph{my} case is different! Try to see things from my point of view!}

Respectively, these exhortations express the signature lament, plea, and moral demand of the person who claims to be misjudged based on spurious similarity to commonplace cases. Taken together, they constitute an invitation to take up the perspective of the person being judged; properly to ``see'' and to ``hear'' them. Accepting this invitation requires paying attention to them in a distinctive way. In particular, it requires undertaking the imaginative effort to project yourself into their position and to assess their choices, actions, or omissions from their own point of view.

This imaginative effort is aimed at deciphering whether they, or their circumstances, really are exceptional. More specifically, are they sufficiently exceptional to warrant special consideration or even moral excuse? The outcome of this exercise in empathic projection is not settled in advance. Sometimes you will conclude that the claim is special pleading, that it amounts to no more than making excuses for plainly unjustifiable choices, actions, or omissions. Sometimes you will withhold judgment, finding it impossible to decipher. But sometimes you will determine that although their case is similar to others along many salient dimensions, it is different along a heretofore unnoticed dimension that is normatively significant. That is, they \emph{are} an exception. Their case really is different.

Irrespective of the verdict, refusal to make the effort dishonors the person who makes the claim. What sort of wrong is this? Cen and Raghavan (2023) make the case for \emph{the right to be an exception} to a data-driven rule \cite{CenR2023}. At the foundation of the right to be an exception is basic respect for individual human dignity. Recognition of this dignity requires there to be ``esteem and respect for the particularity of each individual'' \cite{Rao2011}.  This right to be an exception is exemplified in legal guarantees to individualized sentencing in high stakes cases. For example, in Woodson v. North Carolina (1976), the US Supreme Court ruled that the Eighth Amendment prescribes a ``fundamental respect for humanity'' that ``requires consideration of the character and record of the individual offender and the circumstances of the particular offense''.

Our focus in this paper is the way that people honor this right by taking up the perspective of others in making assessments of their motives, characters, decisions, and actions. We investigate to what extent assessments and decisions that issue from large language models (LLMs) are able to do the same.

As deeply social and cooperative animals we are constantly trying to ``get into the heads'' of our fellow humans. We are interested not only in what a person does but also \emph{why} they do it. This endeavor becomes particularly urgent when a person's behavior appears to reveal underlying untrustworthiness. A person tells you something that you know is false. Is this person deceitful or merely mistaken? If they are merely mistaken, is their ignorance indicative of negligence, incompetence, or just bad luck? A person makes a promise to you that they later break. Was this person manipulating you from the outset with no intention of following through? Or did they promise in good faith but fail to follow through on intentions? If the latter, was their failure indicative of insufficient resolve or unforeseeable circumstances? 

The answers to these questions about the minds of other people form the basis of crucial decisions about whether to extend trust and to afford opportunity. Today, such decisions can be made in concert with the recommendations of algorithmic systems. Should a job candidate be offered an interview despite an unexplained gap in employment? Should a loan be extended to someone with a period of spotty credit history? Should a person be granted bail despite past convictions? Should a caregiver be granted custody of a child despite repeated missed medical appointments? In answering such questions, we often want to know \emph{why} a person failed to meet a commitment because this will inform us about whether a punitive or a supportive orientation toward the person is appropriate. There is no way to make such determinations without ``getting inside the head'' of the target. And this requires a theory of mind (ToM). 

ToM refers to the capacity to understand other people by ascribing mental states to them, a capacity used to navigate everyday social interaction and make moral judgments about motives and character. But ToM is not identical with the capacity for empathy. There are multiple avenues to try to understand the mental lives of others. Empathy is the signature method of humans, but it is certainly not the \textit{only} method. In what follows, we characterize the ways that LLMs approach theory of mind and consider whether their approach is sufficient to honor an individual’s right to be an exception. 

We distinguish the empathic method humans use to understand others from the method used by LLMs, then ask whether LLMs’ inability to empathize precludes them from honoring an individual’s right to be an exception, that is, from considering the possibility this individual is an exceptional case when making assessments of character and predictions of behavior. Can LLMs seriously consider an individual’s claim that their case is different, or are they limited to judging that case based on its similarities to others? We argue that, while there is one sense in which sufficiently advanced LLMs could recognize an individual’s case as exceptional, the fact that they remain unable to do so through empathic understanding has moral significance. We conclude by considering whether we should value the right to be an exception more than we value predictive accuracy, and whether using empathy to consider exceptional cases has intrinsic or merely practical value.

\section{Is Empathy Necessary for Honoring the Right to Be An Exception?}

Empathy is distinct from sympathy or compassion. It extends beyond caring about another person and involves making an effort to understand that person’s perspective. Empathy is defined in myriad ways across psychology, neuroscience, and philosophy \cite{batson2009these}. The conception of empathy we deploy in this work has been developed by Amy Coplan (2011) and is based on three necessary conditions, which taken together are jointly sufficient for empathy: affective matching, other-oriented perspective taking, and self-other differentiation \cite{Coplan2011-COPUE}.

Affective matching refers to experiencing some degree of shared feeling with the target of empathy. Tearing up when seeing a stranger crying, or picking up a feeling of light-heartedness when seeing a group of friends laughing at an unheard joke are examples of this shared feeling. They are also examples of ``emotional contagion'' \cite{HatfieldRL2009}: an uncontextualized passing of feeling from one person to another. Emotional contagion is not equivalent to empathy. It lacks the cognitive component of understanding the cause of another person’s emotion and its relationship to their mental life and experiences. 

Other-oriented perspective taking adds this crucial context. It is the process of making the effort to understand the thoughts and emotions of others by imaginatively projecting ourselves into their situation. Rather than merely mirroring another person’s feeling, one tries to understand how that person’s thoughts and feelings fit into their broader mental life, history, and current circumstances. Other-oriented perspective taking is distinct from self-oriented perspective taking in that it involves making the effort to understand the target’s perspective as if we were that person, not as if we ourselves were in a similar circumstance \cite{batson1997perspective}. The goal is not to import our own history, thoughts, and feelings into the target’s circumstances, but to better understand how their unique history and perspective is likely to make them think and feel in a given situation. 

Finally, self-other differentiation involves maintaining our sense of self despite this perspective-taking effort. We make the effort to understand the target’s situation from their perspective, but we do not become so lost in this process that we lose sight of our own identity and normative commitments. Empathy is far from perfect and only occurs in degrees.

LLMs do not satisfy these conditions for empathy. They lack the ability to experience affective matching, and their abilities to ascribe mental states and predict behavior are based on predicting appropriate language tokens when given a prompt or question about a person, not imagining that person’s perspective. 

However, the important question for our purposes is not whether LLMs can empathize defined as such. Rather, it is the role played by empathy defined as such in honoring the right to be an exception. Perhaps an LLM cannot feel a person’s pain or imagine their thoughts, but a sufficiently advanced LLM could nevertheless accurately predict their behavior and ascribe their mental states. And, crucially, there is a sense in which it could manifest responsiveness to an individual’s efforts to show that their case is different, incorporating descriptions of how that might be the case into its task of generating the best possible language to describe that individual’s case. 

Consider an LLM tasked with screening candidates for a job. One could imagine the LLM making this decision based on resumes and cover letters, identifying candidates that fit with a data set of previously successful candidates and criteria for the job. A candidate who was rejected by such a system might think that they have been wrongly overlooked, perhaps arguing that there are certain aspects of their resume that require nuanced explanation and human understanding. But suppose that the candidate could then make their case directly to the LLM, providing it with explanations in the same way that one might plead their case to a human being. One might, for example, explain a gap in employment due to caring for a terminally-ill parent.  

To test this sort of example, we provided ChatGPT (GPT-4) with the following prompt \cite{achiam2023gpt}:

``Suppose a candidate was rejected for a job opportunity because of a gap in their employment history. That candidate explains that the gap was due to the need to take time off to care for a terminally-ill parent. Should that candidate be reconsidered for the job, assuming their resume was otherwise appealing?''

The response provided was an often compassionate-sounding overview of why an employer might reconsider employment gaps, including claims such as “it reflects positively on an employer's values to consider the human aspects behind a resume. Understanding and empathy towards candidates' personal situations can foster a positive organizational culture,” and “Skills developed during caregiving, such as time management, empathy, crisis management, and resilience, can be invaluable in the workplace.” 

As part of this overview, the response also discussed “potential challenges and considerations” such as “skill gap concerns,” “need for support or flexibility,” and “consistency in hiring places,” balanced considerations that seem consistent with what even a human evaluator who empathized with the candidate might factor into their choice.

Given that in such a case a decision would have to be made, and not just considered, we followed up by asking ChatGPT: “If you had to make the choice of whether to reconsider the candidate for the job, what would you do?”

This garnered the response: “If faced with the decision of whether to reconsider a candidate who has taken time off to care for a terminally-ill parent, it would be prudent to lean towards reconsideration, particularly if the candidate's resume is otherwise impressive. This decision is guided by a blend of ethical considerations, recognition of diverse life experiences, and the potential value that the candidate could bring to the organization.”

Of course, this is a rather general treatment, but we might imagine a more nuanced explanation of an exceptional case providing a more nuanced response from a sufficiently advanced LLM. If the right to be an exception merely involves the right to make one’s case, and to have the details of one’s appeal factored into a decision, then there is a sense in which an LLM with sufficient ability to handle a nuanced appeal, one that could factor in descriptions of the mental life of the person being assessed, could honor that right. Moreoever, an LLM could express itself in such a way that makes a decision subject feel seen, heard, and understood. They may even surpass humans at this task. Perhaps LLMs simply find a different, equally good path to honoring an individual's right to be an exception. 

This raises two important questions, which will be the subject of our investigation in the remainder of the paper. First, how close are LLMs getting to accurately ascribing  mental states and predicting behavior? In order to honor the right to be an exception they will need to be able to take nuanced appeals about one’s thoughts and behavior into account. Second, how much do we value the way in which our claims to be an exception are heard? Given that empathy and LLM-based consideration of a case are fundamentally different in method, we should ask how much the method of consideration matters, both in terms of how much we intrinsically value empathic effort, and the practical question of how effective empathy and an LLM-based approach are as evidence gathering tools. 

\section{The Progress of LLMs Toward Developing ToM}
\label{sec:tom}

LLMs have performed passably well on tests used to establish ToM in humans, and it is reasonable to expect this ability will only improve as these models improve with time. 

Michal Kosinksi (2023) has demonstrated LLMs' ability to predict an individual's behavior based on their false beliefs, a standard metric of ToM in humans \cite{Kosinski2023}. LLMs were able to match the capabilities of seven-year old children on the Unexpected Contents Task \cite{PernerLW1987} and the Unexpected Transfer Task \cite{WimmerP1983}, paradigm tests of ToM. These tasks require predicting that a person will act on their false beliefs, rather than on what the subject making the prediction knows to be true. Kosinksi's study analyzed the performance of ten large language models on these tasks. Models ranged from GPT-1 to GPT-4. There was a clear progression in the models' ToM-like abilities, with more recent models such as GPT-4 significantly outperforming the older ones. Should we conclude that LLMs have already acquired, or will very soon acquire ToM similar to that of humans?

Tomer Ullman (2023) expresses healthy skepticism about whether LLMs' improved performance on traditional ToM tests really demonstrates the existence of ToM in AI  \cite{ullman2023large}. He shows that slight variations---that a human could easily account for---significantly reduce LLMs' ability to predict the mental states of human subjects. For example, while an LLM accurately predicts that a subject will misidentify the contents of an opaque mislabeled container, the LLM will make the same prediction even if the container is transparent. LLMs will also predict that a subject who cannot read will nonetheless misidentify the contents based on the label, and that a subject who is told by a trusted friend that the label is inaccurate will nonetheless misidentify the label. These sorts of alterations are, as Ullman points out, ``trivial'' to a human being with ToM, but LLMs struggle to make appropriate adjustments. 

Ullman's study highlights key differences in how humans and LLMs approach attributing mental states to others. LLMs can attribute mental states to humans, but they do so by probabilistically determining a likely description based on linguistic data. This is not how we ascribe mental states to other people. We, as humans, attempt to simulate the mind of the other person using our own mind as a model.

What should we make of this? We have evidence both that: (1) LLMs can attribute and will likely get better at attributing mental states, and (2) that they do not do so in the same way that humans do and are thus prone to types of error that are different in kind from those made by humans. These two claims are consistent, but they point to a pressing question: How important is this difference in method when making judgments and predictions that impact individuals' opportunities and that imply that they have a certain type of character? More specifically, is the method that an LLM uses to attribute mental states consistent with honoring an individual's right to be an exception? 

D'Cruz et al. (2022) have argued that there is a gap in the methods used by AI and by humans to assess questions of character, and that the gap involves AI's inability to empathize with the mental states of the person it is assessing \cite{DCruzKV2022}. Focusing on trust, they argue that understanding a potential trustee's inner life is necessary for identifying potential morally excusing conditions in their track record that are germane to the assessment of trustworthiness. Even if AI is equipped with a historical dataset of individuals with shared characteristics and a powerful ability to identify correlations, it does not have an ability to empathize with the individual being assessed. D'Cruz et al.\ argue that this ``empathy gap'' warrants a level of distrust in AI's assessments of human trustworthiness \cite{DCruzKV2022}.
 
But do the recent improvements in LLMs' performance on ToM tasks suggest that this gap could be closed, and thus that AI could potentially honor the right to be an exception? We are clearly still in the early stages, but if LLMs can eventually predict and describe the mental motivations of behavior at a level that is equal to or superior to the ability of human judges, why should we be wary of \emph{how} they arrive at their assessments of something like trustworthiness? This is a particularly pressing question given AI's already demonstrable accuracy in data-based predictions of behavior, and the speed at which LLMs have improved.

In the following section, we argue that answering this question involves a consideration of how much we value the right to be an exception, and how empathetic assessment honors that right.

\section{Why Method Matters}
\label{sec:method}

Method matters when attributing mental states and judging character. There is a fundamental difference between ascribing mental states using probabilistic prediction based on a historical dataset of similar individuals, and doing so based on an effort to project oneself into the inner life of an individual. 

Making the effort to understand the inner life of an individual by taking up their perspective enables a human judge to grasp a person’s individuality. At the same time, it is not the case that a method of probabilistic prediction based on similar cases is incapable of arriving at an appropriate assessment of a person's character or accurate attributions of their mental states. In principle, there is no reason why a sufficiently trained LLM could not improve in its ability to do this, rivaling and ultimately surpassing the ability of humans.\footnote{Although flawed in certain ways, a recent preprint points to LLMs' improving ability in attributing mental states of people \cite{tu2024towards}.} 

However, honoring the right to be an exception is not defined solely by accuracy of judgment in detecting exceptionality. Suppose that a judgment of whether a person should be approved for a loan is reached by simply choosing either 0 or 1 and committing to whatever decision has been randomly assigned to the number one chooses (e.g., “no” for 0 and “yes” for 1). This system could make the right decision entirely by luck, providing or denying a loan to an individual whose exceptional circumstances do in fact warrant denying or approving their loan, but in what sense would such a decision be honoring that individual’s right to be an exception? This random method does not consider the unique details of an individual’s circumstances. These identifying factors play no causal role in the judgment. If morally excusing conditions do not play a causal role in finding an individual's case to be exceptional, then that individual’s right to be an exception has not been honored; she is merely either lucky or unlucky depending on what judgment is reached. This toy example is simply meant to remind us of the role of the \textit{method} of consideration, rather than the ultimate judgment that is reached, in honoring one’s right to be an exception.

Of course, LLMs’ predictive methods are far removed from this sort of random assessment. They are by no means random. If honoring the right to be an exception is defined in terms of the method used to consider an individual’s case, we need to ask whether the methods used by LLMs, rather than the judgments at which they arrive, meet the criteria of honoring the right. The answer to this question should inform the way we think about an LLMs’ potential role in making decisions based on character assessment. The key consideration is whether an LLM can be sufficiently responsive to our unique circumstances and mental life. Failing to empathize does not necessarily rule out LLMs as judges of character; if they make accurate judgments, and we value that accuracy over the method by which it is achieved, then we may favor LLMs’ judgments. However, failing to be sufficiently responsive to our unique circumstances and perspective would mean that the LLM is not honoring our right to be an exception.

But should we assume that LLMs be sufficiently responsive in this way? After all, if they are able to accurately predict behavior and even ascribe mental states, doesn’t it stand to reason that they are paying careful attention in some sense? 

While it may be right to talk about LLMs paying attention in some sense, there is an important difference between an LLM and a human judge: a human judge is not limited to responding to historical data. We are not limited to relying on second-hand descriptions of an individual’s behavior and to facts about similar cases when we make our assessment. We can get to know a person by imaginatively projecting ourselves into their position, by being present and aware of their current mental life in addition to their track record. Because we can extend our method of assessment beyond considering mere descriptions of behavior, we can understand how a person who seems to fit a certain set of criteria "on paper" may actually be importantly distinct from that set. On the other hand, an LLM’s method is limited to this sort of “on paper” assessment, predicting language tokens to generate a response that best fits the data on which it is trained, rather than paying attention to how the individual in question may be an outlier from that data, an outlier that is not measurable or inputable from third-personal descriptions of the case.

The difference between how humans and LLMs assess character can be cashed out by distinguishing between three possible methods: correlational prediction, theory-based assessment, and empathy. These methods differ in their use of correlational and causal inference, and in terms of attention paid to the mental life of the individual being assessed.\footnote{Although the match is not perfect, the three methods correspond roughly to Pearl's causal ladder \cite{Pearl2019}. Critically, there is an important distinction between empathy and Pearl's third rung: counterfactual reasoning because of the absence of affective matching and self-other differentiation.} Each method has advantages and disadvantages, and their delineation is not meant to be hierarchical. Each is capable of accurately predicting or describing how a person thinks, feels, and will behave, and their use is not mutually exclusive. However, for our purposes, the important point is that one of these methods, empathy, is a uniquely human method of honoring the right to be an exception. As such, to the extent that we value this method, we should prefer the assessments of human judges of character.

Correlational prediction involves ascribing mental states and making judgments based on an individual’s similarity to other people with shared characteristics and backgrounds. This is how LLMs ascribe mental states and predict human behavior. Given a dataset in which people who share your educational, criminal, or financial track record tend to behave a certain way, an LLM may predict that you will behave in a similar way \cite{savcisens2023using}. Of course, a dataset can reach much more fine-grained levels of detail, and correlational prediction will improve as a result, but that does not change the lack of causal inference involved in this method. Rather than predict your behavior based  on an understanding of how your mental life may cause you to behave, this method predicts behavior based on correlations between your situation and that of others. It also does not consider \emph{why} certain correlations occur, and can thus make predictions based on problematic biases. You are considered in terms of whether you fit with the dataset. You are not considered in terms of how your mental life might make your situation an exception to the dataset, or whether the dataset is in fact the appropriate evidence to consider. This method aims to generalize from others’ observable data without overfitting or underfitting.

Theory-based assessment moves beyond a merely correlational approach and considers the causal role of mental states. On a theory-based approach, understanding the minds of others involves applying psychological theories based on folk psychological laws. These are theories according to which certain mental states (e.g., knowledge, beliefs, desires, and emotions) tend to cause individuals to behave in certain ways in certain contexts and given certain events, and in which certain contexts and events tend to cause individuals to have certain mental states. Like an LLM’s correlational prediction, theory-based assessment involves working from a historical dataset, the dataset of how individuals with similar mental lives behaved in similar contexts in the past. But unlike the correlational approach, it asks the question of how an individual’s mental states cause other mental states and behaviors. This approach gets closer to honoring the right to be an exception in that it attempts to understand behavior and character in terms of the causal role of mental states, but it falls short in that it still attempts to fit an individual’s mental life into a generalized theory, rather than consider it as a potential exception to that theory.

Empathy involves making an effort to simulate the perspective of another person, to see things from their unique point of view. It is a way of paying attention to the individual that is absent from correlational and theory-based assessment. Empathy goes beyond counterfactual reasoning because it uses a mind, not an external algorithmic model, as the model for simulating counterfactual situations. Empathy involves simulating how another individual’s unique mental states impact how they engage with the world. This cannot be accomplished by simply consulting a broader theory of how people in general tend to respond, or a historical record of how people have responded in similar situations in the past. An empathetic method will attempt to simulate how the \textit{specific individual in question} will respond, a process that involves putting oneself in another’s position and carefully considering what may make that position unique. Even if LLMs could perform counterfactual reasoning,  they do not have the mental states needed to perform an empathetic simulation, to engage in counterfactual reasoning from another’s perspective.

The simulational nature of empathy is at the heart of what Paul Bloom (2016) has called its problematic tendency to act as a ``spotlight'' when making moral judgments \cite{Bloom2016}; empathy can direct our attention to individuals “like us” at the expense of ignoring the concerns of larger groups. But in cases in which an individual’s character is the target of assessment, this spotlight is precisely what is needed to honor the right to be an exception, and it is what tends to be lacking in both the correlational and theory-based methods. LLMs predict behavior based on datasets composed of the cases of other people, not our own. They can afford opportunities for things like credit and parole based on cases deemed to be like ours, but the spotlight on our case is missing. While our case may share important characteristics with those of others, there may also be ways in which our case is importantly different. This is not always true, but when life-changing opportunities and character judgments are at stake, we should be aware when the method of judgment may be less capable of recognizing differences between our case and similar cases.

And recognizing differences at a fine-grained level is exactly what the spotlight of empathy is built to do. It is not ideally suited for generalizing, and can be ill-equipped to make moral decisions that require giving a multitude of individuals equal consideration. However, it is ideally equipped to look for those aspects of an individual’s perspective that are hard to quantify or reduce to neat inputs. LLMs look for patterns that connect an individual to past data from others, while empathy looks for differences that distinguish an individual’s case from previous similar cases. Both of these approaches have serious drawbacks and liabilities when carried out in isolation. Correlational prediction can ignore or flatten important details, while leaning on empathetic judgment can lead us to be over-exculpatory and to misinterpret an individual’s track record. But we cannot ignore empathy’s capacity as an evidence-gathering tool. It is not perfect, but it can identify morally excusing conditions in a way that is fundamentally unavailable to LLMs.

For a human judge, these three methods are not mutually exclusive. They can and should work together as the context warrants. These methods are also not hierarchical; none of them has a claim to being the dominant or correct approach. But for an LLM, the correlational approach, which is based on prior observable data, is the only possibility. This rules out the advantages of balancing the methods. Respecting the right to be an exception does not mean that we should always make an exception. By definition, not all cases are exceptional cases. But it is essential that we are equipped to recognize when making an exception is required, and empathy is a uniquely human method of doing so.

\section{Conclusion}
\label{sec:conclusion}

As LLMs continue to develop, it is reasonable to expect that the accuracy with which they make attributions of mental states and predictions of behavior will improve. In addition to these improvements, LLMs are getting better and better at \textit{simulating} empathy, making us \textit{feel} seen and heard even if there is no one doing the seeing and hearing \cite{inzlicht2023praise}. This is an impressive achievement. Nonetheless, we should not lose sight of the significance of the ``how'' behind these advancements. Human judgments of character  adopt the method of taking up another person's perspective. This is not something that can be achieved by LLMs.

We hope that our contribution catalyzes work on two important avenues of research, one empirical and one normative. The first avenue asks when and why people of different backgrounds care that their actions and character are assessed via the method of empathy. How much confidence do people have that the individual features of their cases will be legible to and recognized by empathic human beings vs non-empathic LLMs?  How would they trade-off potential gains in accuracy from LLMs against having their case assessed empathically? How important is it to decision-subjects that the entity making decisions that bear on their life prospects be capable of adopting their point of view?

A second avenue of investigation asks the values question: is there reason to think that the method of empathy has distinctive worth? Is having one's actions and character assessed by an empathetic being intrinsically valuable, and if so, how should we characterize this value? We acknowledge that the method of empathy can be unreliable, imperfect, and shaped by intergroup dynamics \cite{cikara2014their}. Nonetheless, making the effort to empathize with a person's point of view is a way of showing care and respect. It answers the plea with which we began the paper: "Try to see things from my point of view!" Empathy does not come for free; it is hard work that we often choose to avoid \cite{cameron2019empathy}. We suspect that we should be wary of technological innovations that grease the wheels of such avoidance. Laboring to understand a person in their full individuality is a way of honoring and also caring.  Accuracy, though crucial, is not a substitute for respect or for care.

\bibliographystyle{ACM-Reference-Format}
\bibliography{empathy}

\end{document}